\documentstyle[12pt]{article}
\def\seCtion#1{\section{#1} \setcounter{equation}{0}}
\renewcommand\theequation{\ifnum\value{section}>0{\thesection.
\arabic{equation}}\fi}
\newcommand{\be}{\begin{equation}}
\newcommand{\ee}{\end{equation}}
\newcommand{\bea}{\begin{eqnarray}}
\newcommand{\eea}{\end{eqnarray}}
\newcommand{\nn}{\nonumber}
\newcommand{\bi}{\bibitem}
\newcommand{\la}{\label}

\begin{document}
\pagestyle{empty}
\begin{flushright}
\mbox{FIUN-CP-99/1}
\end{flushright}
\quad\\
\begin{center}
\Large {\bf Fermionic dispersion relations at finite \\
temperature and non-vanishing chemical \\
potentials in the minimal standard model}
\end{center}
\begin{center}
\vspace{0.8cm}
{\bf J. Morales \footnote{Associate researcher of Centro Internacional
de F\'{\i}sica, Santaf\'e de Bogot\'a, Colombia. E-mail
johnmo@ciencias.ciencias.unal.edu.co}, C. Quimbay
\footnote{Associate researcher of Centro Internacional de F\'{\i}sica,
Santaf\'e de Bogot\'a, Colombia. E-mail
carloqui@ciencias.ciencias.unal.edu.co}, and F. Fonseca}\\
{\it {Departamento de F\'{\i}sica}, Universidad Nacional de Colombia\\
Ciudad Universitaria, Santaf\'e de Bogot\'a, D.C., Colombia}\\
\vspace{0.8cm}
June 1, 1999

\end{center}

\vspace{1.7cm}

\begin{abstract}
We calculate the fermionic dispersion relations in the minimal
standard model at finite temperature in presence of non-vanishing
chemical potentials due to the CP-asymmetric fermionic background.
The dispersion relations are calculated for a vacuum expectation
value of the Higgs field equal to zero (unbroken electroweak
symmetry). The calculation is performed in the real time formalism
of the thermal field theory at one-loop order in a general $\xi$
gauge. The fermionic self-energy is calculated at leading order in
temperature and chemical potential and this fact permits us to
obtain gauge invariant analytical expressions for the dispersion
relations.

\vspace{0.8cm}


\end{abstract}

\newpage

\pagestyle{plain}


\seCtion{Introduction}

\hspace{3.0mm}
Interest in the knowledge of propagation properties of relativistic
fermions in plasmas at finite temperature has increased during the last
years. It is well known that the interaction of a fermion with a plasma
in thermal equilibrium at temperature $T$ modifies the Fermionic
Dispersion Relations (FDR) with respect to the zero temperature situation.
This phenomenon has been extensively investigated for the non-dense plasma
case \cite{kalas}-\cite{rio}, i.e. when chemical potential $(\mu_f)$
associated to the fermions of the thermal plasma is equal to zero:
$\mu_f = 0$ and $T \not = 0$. In this case, the FDR have been studied both
for massless fermions in \cite{kalas}-\cite{gat} and for massive fermions
in \cite{pis}-\cite{rio} and they only are gauge invariant if the leading
temperature terms of the one-loop self-energy are taken
\cite{kob,fre,wong,qui}. The FDR describe the propagation of the fermionic
excitations of the plasma (quasi-fermions and quasi-holes) through the
thermal background. These excitations are originated in the collective
behaviour of the plasma system at low momentum. The thermal interaction
of a fermion with the plasma is reflected in the existence of effective
masses for quasi-fermions $(M_f)$ and for quasi-holes $(M_h)$. For
massless fermions, it is known that $M_f = M_h$, whereas for massive
fermions it is observed that $M_f \not = M_h$ \cite{pet, qui}. For the
non-density case, the effective masses are called thermal effective
masses due to $M_{f,h} \sim gT$, being $g$ the interaction coupling
constant.

On the other hand, dispersion relations describing the propagation of
the fermionic excitations of a dense plasma at finite temperature can be
found in the literature \cite{kaj}-\cite{per}. For the dense plasma case
at finite temperature, i.e. $\mu_f \not = 0$ and $T \not = 0$, the FDR
have been calculated both for massless fermions in \cite{kalas, lev, bla}
and for massive fermions in \cite{kaj}-\cite{kal}. In particular,
neutrinos traversing a dense medium have been a subject of great interest
\cite{man}-\cite{per}. As it is expected, the interaction of a massless
fermion with the dense plasma modifies the FDR with respect to the
non-dense plasma case. The effective mass of the quasi-particles $(M_E)$
now satisfies $M_E \sim gT$ and $M_E \sim g\mu_f$, being this the source
of the difference between the FDR in both cases.

The one-loop fermionic dispersion relations at finite temperature and
vanishing chemical potential in the Minimal Standard Model(MSM) were
presented in detail in \cite{qui}. In that work, gauge invariant
analytical expressions for the FDR were obtained in the case when the
vacuum expectation value of the Higgs field equal to zero (unbroken
electroweak phase).

In this paper, working too in the MSM before the electroweak spontaneous
simmetry breaking, we consider the propagation of the fermionic
excitations of a dense plasma at finite temperature. The dense plasma is
characterized by non-vanishing chemical potentials associated to the
different quark and lepton flavours. The chemical potentials due to the
CP-asymmetric dense plasma, which is generated by the CP-violation of the
electroweak interaction through the Cabbibo-Kobayashi-Maskawa (CKM)
\cite{cabi} matrix. The observed CP-violation in the decay of neutral
kaons \cite{chris} and more recently the tentative evidence for CP
violation in the decays of neutral $B$ mesons observed in the CDF
experiment at Fermilab, it permits us to think that the CP symmetry is an
approximate symmetry for the electroweak interaction \cite{nir}. The main
goal of this paper is to calculate the FDR at finite temperature and
non-vanishing chemical potentials for the different fermion sectors of
the MSM. The calculation is performed in the unbroken electroweak phase
at one-loop order in a general $\xi$ gauge. We carried out the calculation
in the real time formalism of the thermal field theory
\cite{dol}-\cite{land}. We will prove that FDR are only gauge independent
at leading order in temperature and chemical potential.

In section 2, working in the framework of a non-abelian gauge theory where
particles are massless, we calculate the FDR at finite temperature and
non-vanishing chemical potential. The calculation is performed at one-loop
order, initially in the Feynman gauge. We obtain generic expressions for
the Lorentz invariant functions that can be used in particular cases. The
analytic expressions for the FDR are calculated at leading order in
temperature and chemical potential. In section 3, using the results
obtained in section 2, we calculate the FDR for the different fermion
sectors of the MSM case. We consider that the thermal dense plasma is
characterized by $\mu_{f_i} \not = 0$, where $f_i$ represents the
different fermion flavours of the MSM. In section 4, we calculate the
Lorentz-invariant functions in a general $\xi$ gauge (being the Feynman
gauge, the case when $\xi = 0$). We will prove that the leading terms are
gauge independent and this justifies why in section 2, we only take these
terms to calculate the Lorentz-invariant functions. Finally, our
conclusions are summarized in the last section.


\seCtion{FDR in a non-abelian gauge theory}

\hspace{3.0mm}
In this section, we calculate the FDR in the framework of a non-abelian
gauge theory where fermions, gauge bosons and scalars are massless.
The calculation is performed in the real time formalism of the thermal
field theory in the Feynman gauge. The real part of the fermionic
self-energy is evaluated at one-loop order, considering only their finite
temperature and chemical potential leading contributions. We consider that
massless fermions propagate through a dense plasma at finite temperature.
This thermal plasma is characterized by $\mu_{f_i} \not = 0$, where $f_i$
represents the different fermion species of the theory. In this section,
we will take $\mu_{f_1}= \mu_{f_2}= \ldots = \mu_f$.

At finite temperature and density, the Feynman rules for vertices are the
same as those at $T=0$ and $\mu_f=0$, while the propagators in the Feynman
gauge for massless gauge bosons $D_{\mu \nu}(p)$, massless scalars $D(p)$
and massless fermions $S(p)$ are \cite{kobes}:
\bea
D_{\mu \nu}(p) &=& -g_{\mu \nu} \left[ \frac{1}{p^2+i\epsilon} -i
\Gamma_b(p) \right],  \la{bp} \\
D(p) &=& \frac{1}{p^2+i\epsilon}-i{\Gamma}_b(p),  \la{ep} \\
S(p) &=& \frac{p{\hspace{-1.9mm}\slash}}{p^2+i \epsilon}+
i p{\hspace{-1.9mm} \slash}{\Gamma}_f(p),  \la{fp}
\eea
where $p$ is the particle four-momentum and the plasma temperature $T$
is introduced through the functions $\Gamma_b(p)$ and $\Gamma_f(p)$,
which are given by
\bea
\Gamma_b (p)= 2\pi \delta(p^2)n_b (p),  \la{db} \\
\Gamma_f (p)= 2\pi \delta(p^2)n_f (p),  \la{df}
\eea
with 
\bea
n_b (p) &=& \frac{1}{e^{(p\cdot u)/T}-1}, \la{nb}\\
n_f(p) &=& \theta(p\cdot u)n_{f}^{-}(p)+\theta(-p\cdot u)n_{f}^{+}(p),
\la{nf}
\eea
being $n_b(p)$ the Bose-Einstein distribution function. The Fermi-Dirac
ditribution functions for fermions $n_{f}^{-}(p)$ and for anti-fermions
$n_{f}^{+}(p)$ are:
\bea
n_{f}^{\mp}(p)= \frac{1}{e^{(p\cdot u \mp \mu_f)/T}+1}.
\eea
In the distribution functions $(\ref{nb})$ and $(\ref{nf})$, $u^{\alpha}$
is the four-velocity of the center-mass frame of the dense plasma, with
$u^\alpha u_\alpha =1$. For a non-abelian gauge theory with parity and
chirality conservation, the real part of the self-energy for a massless
fermion is written as:
\be
\mbox{Re}\,\Sigma^{\prime}(K)=- aK{\hspace{-3.1mm}\slash}-b
u{\hspace{-2.1mm} \slash},  \la{tse}
\ee
being $a$ and $b$ the Lorentz-invariant functions and $K^{\alpha}$ the
fermion momentum. These functions depend on the Lorentz scalars $\omega$
and $k$ defined by {\hspace {0.1 cm}} $\omega\equiv(K\cdot u)$ and
$k\equiv[(K\cdot u)^2-K^2]^{1/2}$. Taking by convenience
$u^\alpha=(1,0,0,0)$, we have $K^2 =w^2-k^2$ and then, $w$ and $k$ can
be interpreted as the energy and three-momentum, respectively. Beginning
with $(\ref{tse})$, it is possible to write:
\bea
a(w,k) &=& \frac{1}{4k^2} \left[ Tr(K{\hspace{-3.1mm}\slash}\,\mbox{Re}\,
\Sigma^{\prime})- w Tr(u{\hspace{-2.1mm}\slash}\,\mbox{Re}\,
\Sigma^{\prime}) \right],  \la{lifa} \\
b(w,k) &=& \frac{1}{4k^2} \left[ (w^2-k^2)Tr(u{\hspace{-2.1mm}\slash}\,
\mbox{Re}\,\Sigma^{\prime})- w Tr(K{\hspace{-3.1mm}\slash}\,\mbox{Re}
\,\Sigma^{\prime}) \right].  \la{lifb}
\eea

The full fermion propagator including only the mass correction is given
by 
\be
S(p)=\frac 1{K{\hspace{-3.1mm}\slash}-\mbox{Re}\,\Sigma ^{\prime }(K)}=
\frac{(1+a)K{\hspace{-3.1mm}\slash}+bu{\hspace{-2.1mm}\slash}}{(1+a)^2
K^2+2(1+a)bK\cdot u+b^2},
\ee
and the propagator poles can be found when: 
\be
(1+a)^2K^2+2b(1+a)(K\cdot u)+b^2=0.  \la{fdr0}
\ee
If Eq.$(\ref{fdr0})$ is writting in terms of $w$ and $k$, then:
\bea
\left[ \lbrack 1+a(w,k)]w+b(w,k)\right] ^2=\left[ [1+a(w,k)]k\right] ^2,
\la{fdr}
\eea
and the FDR can be obtained from Eq.$(\ref{fdr})$. To solve this equation,
first it is required to calculate $a(w,k)$ and $b(w,k)$. These functions
can be calculated from the relations $(\ref{lifa})$ and $(\ref{lifb})$ in
terms of the real part of the fermionic self-energy. The one-loop diagrams
that contribute to the fermionic self-energy are shown in Fig.$(1)$. The
contribution to the fermionic self-energy from the gauge boson diagram
shown in Fig.(1a) is given by
\be
\Sigma (K)=ig^2C(R)\int \frac{d^4p}{(2\pi )^4}D_{\mu \nu }(p){\gamma }^\mu
S(p+K){\gamma }^\nu ,  \la{fse}
\ee
where $g$ is the interaction coupling constant and $C(R)$ is the quadratic
Casimir invariant of the re-presentation defined by $(L^AL^A)_{mn}=C(R)
\delta _{mn}$.

Substituting $(\ref{bp})$ and $(\ref{fp})$ into $(\ref{fse})$, the
fermionic self-energy can be written as
$\Sigma(K)=\Sigma(0)+\Sigma^{\prime}(K)$, where $\Sigma(0)$ is the
zero-temperature and zero-density contribution and $\Sigma^{\prime}(K)$
is the finite-temperature and chemical potential contribution. It is easy
to see that:
\bea
\Sigma(0)=-ig^2C(R) \int \frac{d^4 p}{(2\pi)^4} \frac{g_{\mu \nu}}{p^2}
\gamma^{\mu} \frac{p{\hspace{-1.9mm}\slash}+ K{\hspace{-3.1mm}\slash}}
{(p+K)^2} \gamma^{\nu}
\eea
and 
\bea
\Sigma^{\prime}(K)=2g^2 C(R) \int \frac{d^4 p}{(2\pi)^4}
(p{\hspace{-1.9mm}\slash}+ K{\hspace{-3.1mm}\slash}) \left[
\frac{\Gamma_b(p)}{(p+K)^2}-\frac{\Gamma_f(p+K)}{p^2}+i \Gamma_b(p)
\Gamma_f(p) \right].
\eea
Keeping only the real part $(\mbox{Re}\,\Sigma^{\prime}(K))$ of the
temperature and chemical potential contribution, we obtain:
\be
\mbox{Re}\,\Sigma^{\prime}(K)=2g^2C(R) \int \frac{d^4 p}{(2\pi)^4} \left[
(p{\hspace{-1.9mm}\slash}+ K{\hspace{-3.1mm}\slash}) \Gamma_b(p) +
p{\hspace{-1.9mm}\slash} \Gamma_f(p) \right] \frac{1}{(p+K)^2}.  \la{rse}
\ee
If we multiply $(\ref{rse})$ by either $K{\hspace{-3.1mm}\slash}$ or
$u{\hspace{-2.1mm}\slash}$, take the trace, perform the integrations over
$p_0$ and the two angular variables, we obtain the following integrals
over the modulus of the three-momentum $p=|\vec p|$:
\bea
\frac{1}{4}Tr(K{\hspace{-3.1mm}\slash}\,\mbox{Re}\,\Sigma^{\prime}) &=&
g^2C(R)\int^\infty_0\frac{dp}{8\pi^2} \left[ \left( 4p +
\frac{w^2-k^2}{2k} L_2^+(p) \right) n_b(p) \right.  \nn \\
&+& \left. \left( 2p+ \frac{w^2 -k^2}{2k} \left[ Log \frac{p+w_+} {p+w_-}
-Log\frac{w_+}{w_-} \right] \right) n_f^-(p) \right.  \nn \\
&+& \left. \left( 2p+ \frac{w^2 -k^2}{2k} \left[ -Log \frac{p-w_+}{p-w_-}
+Log\frac{w_+}{w_-} \right] \right) n_f^+(p) \right],  \la{Tk}
\eea
\bea
\frac{1}{4}Tr(u{\hspace{-2.1mm}\slash}\,\mbox{Re}\,\Sigma^{\prime}) &=&
g^2C(R)\int^\infty_0\frac{dp}{8\pi^2} \left[ \left( \frac{p}{k}L_2^-(p)
+ \frac{w}{k}L_2^+(p)\right)n_b(p) \right.  \nn \\
&-& \left. \frac{p}{k} \left( Log \frac{p+w_+}{p+w_-}- Log \frac{w_+}{w_-}
\right) n_f^-(p) \right.  \nn \\
&-& \left. \frac{p}{k} \left(Log \frac{p-w_+}{p-w_-}- Log \frac{w_+}{w_-}
\right) n_f^+(p) \right],  \la{Tu}
\eea
with the logarithmic functions given by 
\be
L_2^{\pm}(p)=Log\left[\frac{w^2-k^2+2wp+2kp}{w^2-k^2+2wp-2kp}\right] \pm
Log\left[\frac{w^2-k^2-2wp+2kp}{w^2-k^2-2wp-2kp}\right],
\ee
where $w_{\pm}=(w \pm k)/2$. These expressions agree with \cite{erd} and
for $\mu=0$ are identical to those derived by Weldon \cite{wel}. The
Lorentz-invariant functions in the Feynman gauge $a^0(w,k)$ and $b^0(w,k)$
are obtained substituting $(\ref{Tk})$ and $(\ref{Tu})$ into
$(\ref{lifa})$ and $(\ref{lifb})$. Following the notation given in
\cite{qui}, by convenience these functions are written as:
\bea
a^0(w,k)=g^2C(R)A^0(w,k,\mu_f), \la{aF0} \\
b^0(w,k)=g^2C(R)B^0(w,k,\mu_f), \la{bF0}
\eea
where the expressions $A^0(w,k,\mu_f)$ and $B^0(w,k,\mu_f)$ are:
\bea
A^0(w,k,\mu_f) &=& \frac{1}{k^2} \int^\infty_0\frac{dp}{8\pi^2} \left[
\left(4p - \frac{wp}{k}L_2^-(p)- \frac{w^2+k^2}{2k}L_2^+(p) \right) n_b(p)
\right.  \nn \\
&+& \left. \left( 2p + \left[ \frac{w^2-k^2}{2k} + \frac{wp}{k} \right]
Log \left[ \frac{w^2-k^2+2wp-2kp}{w^2-k^2+2wp+2kp} \right] \right)
n_f^-(p)\right.  \nn \\
&+& \left. \left( 2p + \left[ \frac{w^2-k^2}{2k} - \frac{wp}{k} \right]
Log \left[ \frac{w^2-k^2-2wp-2kp}{w^2-k^2-2wp+2kp} \right] \right)
n_f^+(p)\right],  \nn  \la{AI} \\
\eea
\bea
B^0(w,k,\mu_f) &=& \frac{1}{k^2} \int^\infty_0\frac{dp}{8\pi^2} \left[
\left(-4wp + \frac{w(w^2-k^2)}{2k}L_2^+(p)+ \frac{p(w^2-k^2)}{k}L_2^-(p)
\right) n_b(p) \right.  \nn \\
&-& \left. \left( 2wp + \left[ \frac{w^2-k^2}{k} \right] \left[p +
\frac{w}{2} \right] Log \left[ \frac{w^2-k^2+2wp-2kp}{w^2-k^2+2wp+2kp}
\right] \right) n_f^-(p)\right.  \nn \\
&-& \left. \left( 2wp - \left[ \frac{w^2-k^2}{k} \right] \left[ p -
\frac{w}{2}\right] Log \left[ \frac{w^2-k^2-2wp-2kp}{w^2-k^2-2wp+2kp}
\right] \right) n_f^+(p)\right].  \nn  \la{BI} \\
\eea
It is possible to write the last expressions as:
\bea
A^0(\omega,k,\mu_f) &=& A(\omega,k,\mu_f)+ A^{non-lead}(\omega,k,\mu_f),
\la{Afg} \\
B^0(\omega,k,\mu_f) &=& B(\omega,k,\mu_f)+ B^{non-lead}(\omega,k,\mu_f),
\la{Bfg}
\eea
where $A(\omega,k,\mu_f)$ and $B(\omega,k,\mu_f)$ are the leading terms,
and $A^{non-lead}(\omega,k,\mu_f)$ and $B^{non-lead}(\omega,k,\mu_f)$ are
the non-leading terms in temperature and chemical potential of the
integrals $(\ref{AI})$ and $ (\ref{BI})$, respectively. As we are
interested in obtaining gauge invariant FDR, then we only keep the
leading terms $A(\omega,k,\mu_f)$ and $B(\omega,k,\mu_f)$ of the integrals
$A^0(w,k,\mu_f)$ and $B^0(w,k,\mu_f)$. The reason of this fact, it will be
explain in section 4. The integrals $A(\omega,k,\mu_f)$ and
$B(\omega,k,\mu_f)$ are:
\bea
A(\omega,k,\mu_f) &=& \frac{1}{k^2}\int^\infty_0\frac{dp}{8\pi^2}
\left[ 2p-\frac{wp}{k}Log \left( \frac{w+k}{w-k} \right) \right]
\left[2n_b(p)+n_f^-(p)+n_f^+(p) \right], \la{Alead} \nn \\
\\
B(\omega,k,\mu_f) &=& \frac{1}{k^2}  \nn \\
&\times& \int^\infty_0\frac{dp}{8\pi^2} \left[ \frac{p(w^2-k^2)}{k} Log
\left( \frac{w+k}{w-k} \right) -2wp \right]
\left[2n_b(p)+n_f^-(p)+n_f^+(p)\right]. \la{Blead} \nn \\
\eea
With the notation given by $(\ref{aF0})$ and $(\ref{bF0})$, the
Lorentz-invariant functions at leading order in temperature and chemical
potential $a^{lead}(w,k)$ and $b^{lead}(w,k)$ can be written as:
\bea
a^{lead}(w,k)=g^2C(R)A(w,k,\mu_f), \la{aolead}\\
b^{lead}(w,k)=g^2C(R)B(w,k,\mu_f). \la{bolead}
\eea
Evaluating the integrals $(\ref{Alead})$ and $(\ref{Blead})$, we obtain
that $a^{lead}(w,k)$ and $b^{lead}(w,k)$ are given by:
\bea
a^{lead}(w,k) &=& \frac{M_E^2}{k^2} \left[ 1-\frac{w}{2k}Log
\frac{w+k}{w-k}\right], \la{a1} \\
b^{lead}(w,k) &=& \frac{M_E^2}{k^2} \left[ \frac{w^2-k^2}{2k}Log
\frac{w+k}{w-k}-w \right], \la{b1}
\eea
where $M_E$ is:
\be
M_E^2=\frac{g^2 C(R)}{8} \left( T^2+\frac{\mu_f^2}{\pi^2} \right) .
\la{em}
\ee
Substituting $(\ref{a1})$ and $(\ref{b1})$ in $(\ref{fdr})$, for the limit
$k<<M_E$, we obtain two solutions: one describing propagation of
quasi-fermions
\be
w(k) = M_E+\frac{k}{3}+\frac{k^2}{3M_E}+{\cal O}(k^3)  \la{dr1}
\ee
and other describing propagation of quasi-holes
\be
w(k) = M_E-\frac{k}{3}+\frac{k^2}{3M_E}+{\cal O}(k^3).  \la{dr2}
\ee
We observe that if $k=0$, $w(k)= M_E$. Then $M_E$ can be interpreted as
the effective mass of the quasi-fermions and quasi-holes. The value of
$M_E$ given by $(\ref{em})$ is in agreement with \cite{kaj,lev,bla}. For
the limit$k>>M_E$, the FDR describing propagation of quasi-particles and
quasi-holesare:
\bea
w(k) &=& k+\frac{M_E}{k}-\frac{M_E^4}{2k^3}Log(\frac{2k^2}{M_E^2})+....
\la{dr3} \\
w(k) &=& k+2ke^{-2k^2/M_E^2}+....  \la{dr4}
\eea
respectively, and for very high momentum the relations $(\ref{dr3})$ and
$(\ref{dr4})$ become the ordinary dispersion relation for a massless
fermion propagating in the vacuum, i.e. $w(k)=k$. We emphasize that the
FDR given by $(\ref{dr1})$-$(\ref{dr4})$ are gauge independent, as it
will be shown in section 4. We note that the contribution to the real part
of the fermionic self-energy from the generic scalar boson diagram shown
in Fig.$(1b)$ has the same form as the gauge boson contribution
$(\ref{rse})$. The factor $2g^2C(R)$ is replaced by $l^2C_{L,R}$, where
$l$ is the Yukawa coupling constant and $C_{L,R}$ is given in terms of the
Clebsch-Gordan coefficients.

It is very easy to demostrate that $\mu_f=\int^\infty_0 dp
\left[n_f^-(p)-n_f^+(p)\right]$. This result leads to the chemical
potential value is associated with the difference between the number of
fermions over anti-fermions. The later means that if a thermal dense
plasma is characterized by $\mu_f > 0$, then exist an excess of
fermions over anti-fermions in the plasma. By this reason, the results
obtained in this section can be interesting for baryogenesis.


\seCtion{FDR in the MSM: electroweak unbroken phase}

\hspace{3.0mm}
In this section, using the results obtained in section 2, we calculate
the FDR for the different fermion sectors of the MSM in the unbroken
electroweak phase. The calculation is performed for a dense plasma
characterized by non-vanishing chemical potentials. We consider for quarks
$\mu_u \not = \mu_d \not = \mu_c \not = ... \not = 0$, for charged leptons
$\mu_e \not = \mu_ \mu \not = \mu_ \tau \not = 0$ and for neutrinos
$\mu_{\nu_e} \not = \mu_{\nu_{\mu}} \not = \mu_{\nu_{\tau}} \not = 0$.
For a non-abelian gauge theory with parity violation and quirality
conservation, the real part of the self-energy for a massless fermion is:
\be
\mbox{Re}\,\Sigma^{\prime}(K)=- K{\hspace{-3.1mm}\slash}(a_{L}L +a_{R}R)-
u{\hspace{-2.2mm}\slash}(b_{L}L +b_{R}R),
\ee
where $L\equiv\frac{1}{2}(1-\gamma_5)$ and
$R\equiv\frac{1}{2}(1+\gamma_5)$ are the left- and right-handed chiral
projectors, respectively. The functions $a_L$, $a_R$, $b_L$ and $b_R$ are
the chiral projections of the Lorentz-invariant functions $a$, $b$ and
they are defined in the following way:
\bea
a &=& a_L L + a_R R, \\
b &=& b_L L + b_R R.
\eea
The inverse fermion propagator is given by 
\be
S^{-1}(K)= {\cal L}{\hspace{-2.5mm}\slash}L+ \Re{\hspace{-2.5mm}\slash}R
\la{ifp}
\ee
where: 
\bea
{\cal L}^{\mu} &=& ( 1 + a_L) K^{\mu} + b_L u^{\mu} \\
{\Re}^{\mu} &=& ( 1 + a_R) K^{\mu} + b_R u^{\mu}
\eea
The fermion propagator follows from the inversion of $(\ref{ifp})$: 
\bea
S=\frac{1}{D}\left[\left({\cal L}^2\Re{\hspace{-2.5mm} \slash}\right)L +
\left(\Re^2{\cal L}{\hspace{-2.5mm}\slash} \right)R \right],  \la{p1}
\eea
being $D(\omega,k)={\cal L}^2 {\Re}^2$. The poles of the propagator
correspond to values $w$ and $k$ for which the determinat $D$ in
(\ref{p1}) vanishes:
\be
{\cal L}^2 {\Re}^2 =0.  \la{d}
\ee
In the rest frame of the dense plasma $u=(1,\vec 0)$, Eq.$(\ref{d})$
leads to the FDR for a chirally invariant gauge theory with parity
violation, as is the case of the MSM before the electroweak
spontaneous simmetry breaking. Thus, the FDR for this case are given by
\bea
\left[ \omega (1+a_L)+b_L \right]^2- k^2 \left[ 1+a_L \right]^2 &=& 0,
\la{dra} \\
\left[ \omega (1+a_R)+b_R \right]^2-k^2 \left[ 1+a_R \right]^2 &=& 0.
\la{drb}
\eea
Left- and right-handed components of the FDR obey decoupled relations.
The Lorentz invariant functions $a$ and $b$ are calculated from the
relations $(\ref{lifa})$ and $(\ref{lifb})$ through the real part of the
fermionic self-energy. This self-energy is obtained additioning all the
posibles gauge boson and scalar boson contributions admited by the MSM
Feynman rules. The diagrams with an exchange of charge Goldstone bosons
or $W$ gauge bosons induce a flavour change in the incoming fermion $I$
to a different outgoing fermion $F$.

For the quark sector, in the case of the flavour change contributions
mentioned, the flavour $i$ of the internal quark (inside the loop) runs
over the up or down quarks flavours according to the type of the external
quark (outside the loop). Since each contribution to the quark self-energy
is proportional to the generic gauge boson contribution $(\ref{rse})$,
then the functions $a_L$, $a_R$, $b_L$ and $b_R$ are given by

\bea
a_{\stackrel{L}{R}}(\omega,k)_{IF} &=& [f_{\stackrel{L}{R}}^{\gamma g}+
f_{\stackrel{L}{R}}^{Z \chi_3}+ f_{\stackrel{L}{R}}^{H}]A(w,k,\mu_{I})
\delta_{IF}+ \sum_{i}f_{\stackrel{L}{R}}^{W \chi}A(w,k,\mu_{i}),  \la{alr}
\\
b_{\stackrel{L}{R}}(\omega,k)_{IF} &=& [f_{\stackrel{L}{R}}^{\gamma g}+
f_{\stackrel{L}{R}}^{Z \chi_3}+ f_{\stackrel{L}{R}}^{H}]B(w,k,\mu_{I})
\delta_{IF}+ \sum_{i}f_{\stackrel{L}{R}}^{W \chi}B(w,k,\mu_{i}).  \la{blr}
\eea
The coefficients $f$ are: 
\bea
f_{\stackrel{L}{R}}^{\gamma g} &=& \frac{4}{3}g_s^2 + g^2 Q_I^2 s_w^2, \\
f_{L}^{Z \chi_3} &=& g^2 \left[ \frac{(T_3 - Q_I s_w^2)^2}{c_w^2}+
\frac{m_I^2}{8m_w^2} \right], \\
f_{R}^{Z \chi_3} &=& g^2 \left[\frac{Q_I^2 s_w^4}{c_w^2}+
\frac{m_I^2}{8m_w^2} \right], \\
f_{\stackrel{L}{R}}^{H} &=& \frac{g^2 m_I^2}{8m_w^2}, \\
f_{L}^{W \chi} &=& \frac{g^2 K_{Ii}^{+}K_{iF}}{2} \left[
\frac{m_i^2}{2m_w^2}+1 \right], \\
f_{R}^{W \chi} &=& \frac{g^2 K_{Ii}^{+}K_{iF} m_I m_i}{4m_w^2},
\eea
where ${\it K}$ represents the CKM matrix, $g_s$ is the strong coupling
constant, $g$ is the weak coupling constant, $Q_I$ is the electric charge
of the incoming quark and $s_w$ ($c_w$) is the sine (cosine) of the
electroweak mixing angle $\theta_w$. The functions $A(w,k,\mu_f)$ and
$B(w,k,\mu_f)$ are given by
\bea
A(w,k,\mu_f) &=& \frac{1}{8 k^2} \left( T^2+ \frac{\mu_f^2}{\pi^2}
\right)\left[1-\frac{w}{2k} Log \frac{w+k}{w-k} \right] ,  \la{ALR} \\
B(w,k,\mu_f) &=& \frac{1}{8 k^2} \left( T^2+ \frac{\mu_f^2}{\pi^2}
\right)\left[\frac{w^2-k^2}{2k} Log \frac{w+k}{w-k}-w \right],  \la{BLR}
\eea
and they have been obtained through the evaluation of the integrals
$(\ref{Alead})$ and $(\ref{Blead})$, which are the non-leading terms in
temperature and chemical potential of the integrals $(\ref{AI})$ and
$(\ref{BI})$, respectively. The chiral projections of the
Lorentz-invariant functions are:
\bea
a_{L}(\omega,k)_{IF} &=& \frac{1}{8k^2}\left[1-F(\frac{w}{k})\right]
\left[l_{IF}(T^2+\frac{\mu_I^2}{\pi^2})+ c_{IF}(T^2+\frac{\mu_i^2}
{\pi^2})\right],  \la{aL1} \\
b_{L}(\omega,k)_{IF} &=& -\frac{1}{8k^2}\left[\frac{w}{k}+(\frac{k}{w}- 
\frac{w}{k})F(\frac{w}{k})\right]\left[l_{IF}(T^2+\frac{\mu_I^2}{\pi^2}) +
c_{IF}(T^2+\frac{\mu_i^2}{\pi^2}) \right],  \nn  \la{bL1} \\
\\
a_{R}(\omega,k)_{IF} &=& \frac{1}{8k^2}\left[1-F(\frac{w}{k})\right]
\left[r_{IF}(T^2+\frac{\mu_I^2}{\pi^2})+ d_{IF}(T^2+\frac{\mu_i^2}
{\pi^2})\right],  \la{aR2} \\
b_{R}(\omega,k)_{IF} &=& -\frac{1}{8k^2}\left[\frac{w}{k}+(\frac{k}{w}- 
\frac{w}{k})F(\frac{w}{k})\right]\left[r_{IF}(T^2+\frac{\mu_I^2}{\pi^2}) +
d_{IF}(T^2+\frac{\mu_i^2}{\pi^2}) \right],  \nn  \la{bR2} \\
\eea
where $F(x)$ is 
\be
F(x)=\frac{x}{2} Log \left(\frac{x+1}{x-1}, \right)
\ee
and the coefficients are given by 
\bea
l_{IF} &=& \left(\frac{4}{3}g_s^2 + g^2Q_I^2s_w^2 + \frac{g^2(T_3-Q_I
s_w^2)^2}{c_w^2} + \frac{g^2m_I^2}{4m_w^2}\right)\delta_{IF}, \\
c_{IF} &=& \sum_i \frac{g^2 K_{Ii}^+ K_{iF}}{2}\left[\frac{m_i^2}
{2m_w^2}+1\right], \\
r_{IF} &=& \left(\frac{4}{3}g_s^2 + g^2Q_I^2s_w^2 +
\frac{g^2Q_I^2 s_w^4}{c_w^2}+\frac{g^2m_I^2}{4m_w^2}\right)\delta_{IF}, \\
d_{IF} &=& \sum_i \frac{g^2 K_{Ii}^+ K_{iF}m_I m_i}{4m_w^2}.
\eea
Substituting $(\ref{aL1})$-$(\ref{bL1})$ into $(\ref{dra})$,
{\hspace{0.1 cm}} and $(\ref{aR2}) $-$(\ref{bR2})$ into $(\ref{drb})$,
{\hspace{0.1 cm}} for the limit $k << M_{E_{L,R}}$, we obtain two
solutions describing propagation of quasi-fermions:
\bea
w(k)_L = M_{E_L}+\frac{k}{3}+\frac{k^2}{3M_{E_L}}+{\cal O}(k^3),
\la{dr1L}\\
w(k)_R = M_{E_R}+\frac{k}{3}+\frac{k^2}{3M_{E_R}}+{\cal O}(k^3),
\la{dr1R}
\eea
and two solutions describing propagation of quasi-holes:
\bea
w(k)_L = M_{E_L}-\frac{k}{3}+\frac{k^2}{3M_{E_L}}+{\cal O}(k^3),
\la{dr2L}\\
w(k)_R = M_{E_R}-\frac{k}{3}+\frac{k^2}{3M_{E_R}}+{\cal O}(k^3),
\la{dr2R}
\eea
where: 
\bea
M_{E_L}^2 &=& T^2(l_{IF}+c_{IF})+l_{IF} \frac{\mu_I^2}{\pi^2}+ c_{IF}
\frac{\mu_i^2}{\pi^2},  \la{emL} \\
M_{E_R}^2 &=& T^2(r_{IF}+d_{IF})+r_{IF} \frac{\mu_I^2}{\pi^2}+ d_{IF}
\frac{\mu_i^2}{\pi^2}.  \la{emR}
\eea

The FDR for the lepton sector are similar to the relations $(\ref{dr1L})$
-$(\ref{dr2R})$, even though the effective masses changing. For this
sector, $g_s = 0$, $Q_I = -1$, $T_3 = -1/2$, and non-exist mixing between
the charge lepton flavours. Then, for the charge lepton sector, the
coefficients $l$, $c $, $r$ and $d$ in $(\ref{emL})$ and $(\ref{emR})$
are:
\bea
l &=& g^2s_w^2 + \frac{g^2(\frac{1}{2}-s_w^2)^2}{c_w^2} +
\frac{g^2m_I^2}{4m_w^2}, \\
c &=& \frac{g^2}{2}\left[\frac{m_i^2}{2m_w^2}+1\right], \\
r &=& g^2s_w^2 + \frac{g^2s_w^4}{c_w^2}+\frac{g^2m_I^2}{4m_w^2}, \\
d &=& 0.
\eea

For the neutrino sector $g_s = 0$, $Q_I = 0$, $T_3 = 1/2$, and non-exist
mixing between the neutrino flavours. Aditionally, because the parity
violation of the electroweak interaction, the coefficients $r$ and $d$
vanishing. The coefficients $l$ and $c$ are given by

\bea
l &=& g^2 \frac{1}{4c_w^2}, \\
c &=& \frac{g^2}{2}\left[\frac{m_i^2}{2m_w^2}+1\right].
\eea


\seCtion{Real part of the self-energy in the $\xi$ gauge}

\hspace{3.0mm}
The gauge invariance of the FDR presented in sections 2 and 3 is studied
in this section. The analytical expresions obtained in
$(\ref{dr1})$-$(\ref{dr4})$ were calculated keeping only the leading
contributions of the integrals $(\ref{AI})$ and $(\ref{BI})$. In this
section, we calculate the real part of the one-loop self-energy
$(\ref{fse})$ in a general $\xi$ gauge. Making this calculation, we can
conclude that the Lorentz-invariant functions $(\ref{a1})$ and
$(\ref{b1})$ allowed to obtain the gauge invariant FDR given by
$(\ref{dr1})$ and $(\ref{dr4})$. We note that the functions $(\ref{a1})$
and $(\ref{b1})$ are $\sim T^2$ and $\sim \mu_f^2$, i.e. they were
obtained with the leading terms of $(\ref{AI})$ and $(\ref{BI})$.

At finite temperature and density, the propagators in the $\xi$ gauge,
both for the massless gauge boson $D_{\mu \nu}^{\xi}(p)$ and massless
scalar boson $D^{\xi}(p)$ are:

\bea
D_{\mu \nu}^{\xi}(p) &=& \left[-g_{\mu\nu}+ \xi \frac{p_{\mu}p_{\nu}}{p^2}
\right]\left[\frac{1} {p^2+i\epsilon}-i{\Gamma}_b(p)\right],  \la{bpxg} \\
D^{\xi}(p) &=& \frac{1}{p^2+i\epsilon}-i{\Gamma}_b(p),  \la{epxg}
\eea
with ${\Gamma}_b(p)$ defined by $(\ref{db})$. The contribution to the
fermionic self-energy from the generic gauge boson diagram of Fig.(1a) is:
\be
\Sigma^{\xi}(K)= ig^2 C(R) \int\frac{d^{4} p}{(2\pi)^4} D_{\mu \nu}^{\xi}
(p){\gamma}^{\mu}S(p+K){\gamma}^{\nu},  \la{sexg}
\ee
where $S(p)$ is given by $(\ref{fp})$. Inserting the expressions
$(\ref{fp})$ and $(\ref{bpxg})$ into $(\ref{sexg})$ and keeping only the
real part $\mbox{Re}\,\Sigma^{\prime \xi}(K)$ of the temperature and
chemical potential contribution to the fermionic self-energy
$\Sigma^{\prime \xi}(K)$, it is possible to write:
\be
\mbox{Re}\,\Sigma^{\prime \xi}(K)= \mbox{Re}\,\Sigma^{\prime}(K) +
\mbox{Re}\,\delta\Sigma^{\xi}(K),
\ee
where $\mbox{Re}\,\Sigma^{\prime}(K)$ is the real part of the temperature
and chemical potential contribution to the fermionic self-energy in the
Feynman gauge, given by $(\ref{rse})$, and
$\mbox{Re}\,\delta\Sigma^{\xi}(K)$ is:
\bea
\mbox{Re}\delta \Sigma^{\xi}(K)=\mbox{Re} \delta_b \Sigma^{\xi}(K) +
\mbox{Re} \delta_f \Sigma^{\xi}(K),
\eea
being 
\bea
\mbox{Re} \delta_b \Sigma^{\xi}(K)=\xi g^2C(R)\int
\frac{d^{4}p}{(2\pi)^4}p{\hspace{-1.9mm}\slash}(p{\hspace{-1.9mm} \slash}
+K{\hspace{-3.1mm}\slash})p{\hspace{-1.9mm}\slash}
\left[\frac{{\Gamma}_b (p)}{p^2(p+K)^2} \right]
\eea
and 
\bea
\mbox{Re} \delta_f \Sigma^{\xi}(K)=-\xi g^2C(R)\int
\frac{d^{4}p}{(2\pi)^4}p{\hspace{-1.9mm}\slash}(p{\hspace{-1.9mm} \slash}
+K{\hspace{-3.1mm}\slash})p{\hspace{-1.9mm}\slash}
\left[\frac{{\Gamma}_f(p+K)}{p^4}\right].
\eea
In the last relations, the denominators are defined by their principal
value. We note that there is a principal-value singularity $1/p^2$ in the
longitudinal boson propagator. Its product with $\delta(p^2)$ is defined
by $-\delta(p^2)/p^2 \rightarrow \delta^{\prime}(p^2)$, where the prime
denotes the derivative with respect to $p_0^2$ \cite{wel}. Using this
definition and following a similar procedure to one developed in section
2, the traces $\frac{1}{4}
Tr(K{\hspace{-3.1mm}\slash}\,\mbox{Re}\,\delta\Sigma^{\xi})$ and
$\frac{1}{4}Tr(u{\hspace{-2.2mm}\slash}\,\mbox{Re}\,\delta\Sigma^{\xi})$
are calculated. With similar expressions to $(\ref{lifa})$ and
$(\ref{lifb})$, the Lorentz-invariant functions in the $\xi$ gauge are
given by
\bea
a^{\xi}(\omega,k) &=& g^2C(R) A^{\xi}(\omega,k,\mu_f), \\
b^{\xi}(\omega,k) &=& g^2C(R) B^{\xi}(\omega,k,\mu_f),
\eea
where 
\bea
A^{\xi}(\omega,k,\mu_f) &=& A^0(\omega,k,\mu_f)-\xi A^{\delta}
(\omega,k,\mu_f),  \la{IAx} \\
B^{\xi}(\omega,k,\mu_f) &=& B^0(\omega,k,\mu_f)-\xi B^{\delta}
(\omega,k,\mu_f),  \la{IBx}
\eea
being $A^0(\omega,k,\mu_f)$ and $B^0(\omega,k,\mu_f)$ the integrals in
theFeynman gauge given by $(\ref{AI})$ and $(\ref{BI})$. As it is shown
in $(\ref{Afg})$ and $(\ref{Bfg})$, these integrals can be written as
$A^0(\omega,k,\mu_f)= A(\omega,k,\mu_f)+ A^{non-lead}(\omega,k,\mu_f)$ and
$B^0(\omega,k,\mu_f)= B(\omega,k,\mu_f)+ B^{non-lead}(\omega,k,\mu_f)$,
where $A(\omega,k,\mu_f)$ and $B(\omega,k,\mu_f)$ are the leading terms
in temperature and chemical potential of the integrals $(\ref{Afg})$ and
$(\ref{Bfg})$. The non-leading terms $A^{non-lead}(\omega,k,\mu_f)$ and
$B^{non-lead}(\omega,k,\mu_f)$ are:

\bea
A^{non-lead}(\omega,k,\mu_f)= \frac{1}{k^2} \int^\infty_0\frac{dp}{4\pi^2}
\left( \left[-\frac{\omega^2+k^2}{4k}L_2^+(p) -\frac{\omega p}{2k}L_2^-(p)
\right]n_b(p) \right.  \nn \\
- \left. \left[ \frac{\omega^2-k^2+2 \omega p}{4k} Log
\frac{p+\omega_+}{p+\omega_-}-\frac{\omega^2-k^2}{4k}Log
\frac{\omega_+}{\omega_-} \right]n_f^-(p) \right.  \nn \\
-\left. \left[\frac{\omega^2-k^2-2 \omega p}{4k} Log
\frac{p-\omega_+}{p-\omega_-}- \frac{\omega^2-k^2}{4k}Log
\frac{\omega_+}{\omega_-} \right]n_f^+(p) \right)  \nn \\
\eea
and 
\bea
B^{non-lead}(\omega,k,\mu_f)= \frac{1}{k^2} \int^\infty_0
\frac{dp}{4\pi^2}\left( \left[-\frac{p(\omega^2-k^2)}{2k}L_2^- (p)+
\frac{\omega(\omega^2-k^2)}{4k}L_2^+(p) \right]n_b(p) \right.  \nn \\
- \left. \left[ \frac{\omega^2-k^2}{4k} \left(2p+\omega \right) Log
\frac{p+\omega_+}{p+\omega_-}-\frac{\omega(\omega^2-k^2)}{4k}Log
\frac{\omega_+}{\omega_-} \right] n_f^-(p) \right.  \nn \\
-\left. \left[\frac{\omega^2-k^2}{4k} \left(2p-\omega \right) Log
\frac{p-\omega_+}{p-\omega_-}+ \frac{\omega(\omega^2-k^2)}{4k}Log
\frac{\omega_+}{\omega_-} \right] n_f^+(p) \right).  \nn \\
\eea

On the other hand, in the integrals $(\ref{IAx})$ and $(\ref{IBx})$, the
$\xi$ gauge dependent terms are given by $A^{\delta}(w,k,\mu_f)$ and
$B^{\delta}(w,k,\mu_f)$. These integrals are:
\bea
A^{\delta}(w,k,\mu_f)=\int^\infty_0\frac{dp}{4\pi^2} \frac{1}{k^2} \left(
\left[\frac{\omega^2-k^2}{2p} +\frac{\omega^2-k^2}{2T}e^{P/T}n_b(p)
\left(1+ 2 \frac{\omega}{k}L_2^{-}(p)\right) \right. \right.  \nn \\
- \left. \left. \frac{1}{2}\left(\omega R^{-}(p)+pR^{+}(p)\right) -\omega
p\left(\omega T^{+}(p)+pT^{-}(p)\right) \right. \right.  \nn \\
-\left. \left.\left(\frac{\omega^2-k^2}{8k} -\frac{(\omega^2-k^2)^2}{16kp}
\left[\frac{1}{p} +\frac{1}{T}e^{p/T}n_B(p)\right]\right)L_2^{+}(p)\right]
n_b(p) \right.  \nn \\
+\left.\left[\left( \frac{\omega^2-k^2+2wp}{4k}Log \frac{p+ \omega_+} {p+
\omega_-}-\frac{\omega^2-k^2}{4k}Log\frac{\omega_+}{\omega_-}
\right)n_f^-(p) \right. \right.  \nn \\
- \left. \left. \left( \frac{\omega^2-k^2-2wp} {4k}Log
\frac{p- \omega_+}{p- \omega_-}-\frac{\omega^2-k^2}{4k}Log
\frac{\omega_+}{\omega_-}\right)n_f^+(p) \right] \right)  \nn \la{IAxg} \\
\eea
and 
\bea
B^{\delta}(w,k)=-\int^\infty_0\frac{dp}{4\pi^2} \frac{1}{k^2}\left(\left[
\frac{\omega (\omega^2-k^2)}{2p}+\frac{\omega^2-k^2}{2T}e^{P/T}n_b(p)
\left(\omega+2 \frac{(\omega^2-k^2)}{k}L_2^{-}(p)\right) \right. \right.
\nn \\
- \left. \left. \frac{\omega}{2}\left(\omega R^{-}(p)+pR^{+}(p)\right)
-(\omega^2-k^2) p\left(\omega T^{+}(p)+pT^{-}(p)\right) \right. \right. 
\nn \\
-\left. \left. \left(\frac{\omega(\omega^2-k^2)}{8k}-
\frac{\omega(\omega^2-k^2)^2}{16kp}\left[\frac{1}{p}+ \frac{1}{T}
e^{p/T}n_b(p)\right]\right)L_2^{+}(p)\right]n_B(p) \right.  \nn \\
+\left.\left[\left( \frac{\omega^2-k^2} {4k}\left(2p+\omega \right) Log 
\frac{p+ \omega_+}{p+ \omega_-}- \frac{\omega(\omega^2-k^2)}{4k}Log
\frac{\omega_+}{\omega_-} \right)n_f^-(p) \right. \right.  \nn \\
+ \left. \left. \left( \frac{\omega^2-k^2} {4k} \left(2p-\omega \right)
Log  \frac{p- \omega_+}{p- \omega_-}+ \frac{\omega(\omega^2-k^2)}{4k}Log
\frac{\omega_+}{\omega_-} \right)n_f^+(p) \right] \right),  \nn
\la{IBxg} \\
\eea
being the functions $R^{\pm}(p)$ and $T^{\pm}(p)$ defined by 
\bea
R^{\pm}(p) &=& \frac{(\omega^2-k^2)^2}{(\omega^2-k^2+2\omega p)^2- 4k^2
p^2}\pm\frac{(\omega^2-k^2)^2}{(\omega^2-k^2- 2\omega p)^2- 4k^2p^2}, \\
T^{\pm}(p) &=& \frac{(\omega^2-k^2)}{(\omega^2-k^2+ 2\omega p)^2- 4k^2
p^2}\pm\frac{(\omega^2-k^2)}{(\omega^2-k^2- 2\omega p)^2-4k^2p^2}.
\eea

We have proved numerically and analitically, in the $k \rightarrow 0$
limit, that:

\bea
A^{\delta}(\omega,k,\mu_f) &=& A^{non-lead}(\omega,k,\mu_f), \\
B^{\delta}(\omega,k,\mu_f) &=& B^{non-lead}(\omega,k,\mu_f).
\eea
The last fact means that substituting $(\ref{Afg})$ and $(\ref{Bfg})$ in
$(\ref{IAx})$ and $(\ref{IBx})$, we can write:
\bea
A^{\xi}(\omega,k,\mu_f) &=& A^{lead}(\omega,k,\mu_f)+
(1-\xi)A^{non-lead}(\omega,k,\mu_f), \la{Axf} \\
B^{\xi}(\omega,k,\mu_f) &=& B^{lead}(\omega,k,\mu_f)+
(1-\xi)B^{non-lead}(\omega,k,\mu_f). \la{Bxf}
\eea
In particulary, we have the Feynman gauge for $\xi =0$ and the expressions
$(\ref{Afg})$ and $(\ref{Bfg})$ are obtained from the expressions
$(\ref{Axf})$ and $(\ref{Bxf})$. For the Landau gauge, $\xi =1$, we have:
\bea
A^{\xi =1}(\omega,k,\mu_f) &=& A^{lead}(\omega,k,\mu_f), \\
B^{\xi =1}(\omega,k,\mu_f) &=& B^{lead}(\omega,k,\mu_f).
\eea
We observe that the analytical expressions for the FDR calculated in
section 2 and 3 are gauge invariant beacuse they were obtained beginning
with the gauge independent integrals $(\ref{ALR})$ and $(\ref{BLR})$.


\section{Conclusions}

\hspace{3.0mm}
We have obtained the FDR at finite temperature and non-vanishing chemical
potential for the different fermion sectors of the MSM in the electroweak
unbroken phase. The calculation was performed at one-loop order in the
real time formalism of the thermal field theory in a general $\xi$ gauge.
We have proven that the analitycal expressions obtained for the FDR are
gauge independent, since we have made the calculation at leading order
in temperature and chemical potential. The dispersion relations obtained
depend of the chemical potentials associated to the different quark and
lepton flavours. We have assumed that $\mu_{f_i} \not = 0$ for all the
$f_i$ quark and lepton flavours. In our calculation, the values of the
differents $\mu_{f_i}$ are unknown. We argue that the chemical potentials
due to the CP-asymmetric dense plasma, which is generated by the
CP-violation of the electroweak interaction through the CKM matrix.
Since $\mu_f \not = 0$ means that it exists a difference between the
number of fermions over anti-fermions in the thermal plasma, we think
that this work can be interesting for baryogenesis.

\section*{Acknowledgments}

This work was supported by COLCIENCIAS (Colombia). We thank Belen Gavela
and Silvia Vargas for their help in the bases of this paper.


\newpage

\begin{figure}[1]
\let\picnaturalsize=N
\def\picsize{3in}
\def\picfilename{Fig1.eps}
\caption{Generic gauge and scalar boson diagrams}
\label{Fig.(1)}
\end{figure}

\end{document}